\documentclass[twoside]{article}
\usepackage{fleqn,espcrc2}

\usepackage{epsfig}
\usepackage[figuresright]{rotating}

\newcommand{\AmS}{{\protect\the\textfont2
  A\kern-.1667em\lower.5ex\hbox{M}\kern-.125emS}}

\title{The Lake Baikal neutrino experiment}

\author{
V.A.Balkanov\address{ Institute for Nuclear Research,
Moscow, Russia \\
$^b$ Irkutsk State University,
Irkutsk, Russia\\
$^c$ Institute of Nuclear Physics, MSU,
Moscow, Russia\\
$^d$ Nizhni  Novgorod  State  Technical University,
Nizhni  Novgorod, Russia\\
$^e$ St.Petersburg State  Marine Technical  University,
St.Petersburg, Russia\\
$^f$  Kurchatov Institute,
Moscow, Russia\\
$^g$ Joint Institute for Nuclear Research,
Dubna,Russia\\
$^h$ DESY-Zeuthen,
Zeuthen, Germany\\
$^i$ KFKI, Budapest, Hungary \\
\vspace {4mm}
presented by G.V.Domogatsky},
I.A.Belolaptikov$^g$, L.B.Bezrukov$^a$,
N.M.Budnev$^b$, A.G.Chensky$^b$, I.A.Danilchenko$^a$,
Zh.-A.M.Djilkibaev$^a$,
G.V.Domogatsky$^a$, A.A.Doroshenko$^a$, S.V.Fialkovsky$^d$,
O.N.Gaponenko$^a$, O.A.Gress$^b$, D.D.Kiss$^i$, 
A.M.Klabukov$^a$, A.I.Klimov$^f$, S.I.Klimushin$^a$,
A.P.Koshechkin$^a$, 
V.F.Kulepov$^d$, L.A.Kuzmichev$^c$, Vy.E.Kuznetzov$^a$,
J.Ljaudenskaite$^b$, B.K.Lubsandorzhiev$^a$,
M.B.Milenin$^d$, R.R.Mirgazov$^b$, N.I.Moseiko$^c$,
V.A.Netikov$^a$, E.A.Osipova$^c$, A.I.Panfilov$^a$, Yu.V.Parfenov$^b$,
L.V.Pankov$^b$,
A.A.Pavlov$^b$, E.N.Pliskovsky$^a$, P.G.Pokhil$^a$, E.G.Popova$^c$,
V.V.Prosin$^c$, A.E.Rzhechitsky$^b$,
M.I.Rozanov$^e$, V.Yu.Rubzov$^b$, Yu.A.Semenei$^b$,
I.A.Sokalski$^a$, CH.Spiering$^h$,
O.Streicher$^h$, B.A.Tarashansky$^b$, T.Thon$^h$, G.Toht$^i$,
R.V.Vasiljev$^a$, R.Wischnewski$^h$, I.V.Yashin$^c$.}

\vspace {-1cm}       
\begin{document}

\begin{abstract}
 We review the present status of the Baikal Neutrino Project
and present 
the results of a search for high energy neutrinos
with the  detector intermediate stage {\it NT-96}.
\vspace{1pc}
\end{abstract}

\maketitle


\vspace {-2cm}
\section{Detector and Site}

The Baikal Neutrino Telescope  is deployed in Lake 
Baikal, Siberia, 
\mbox{3.6 km} from shore at a depth of \mbox{1.1 km}. 
{\it NT-200}, the medium-term goal of the collaboration
\cite{APP}, was put into operation 
at April 6th, 1998 and consists of 192
optical modules (OMs)  -- see fig.1. 
An umbrella-like frame carries  8 strings,
each with 24 pairwise arranged OMs.
Three underwater electrical cables connect the
detector with the shore station. 

In April 1993, the first part of {\it NT-200}, the detector {\it
NT-36} with 36 OMs at 3  strings, was put into operation 
and took data up to March 1995. A 72-OM array, {\it \mbox{NT-72}}, 
run in \mbox{1995-96}. In 1996 it
was replaced by the four-string array {\it NT-96}. 
{\it
  NT-144}, a six-string array with 144 OMs, was taking data
in \mbox{1997-98}.

Summed over 1140
days effective lifetime, $6.6\cdot 10^{8}$ muon events have been 
collected with
\mbox{{\it NT-36, -72, -96, -144, -200}}. 

The OMs are grouped in pairs along the strings. They contain 
37-cm diameter {\it QUASAR} PMs which have been developed
specially for our project \cite{APP,OM2,Project}. The two PMs of a
pair are switched in coincidence in order to suppress background
from bioluminescence and PM noise.
A pair defines a {\it channel}. 
%
%

\begin{figure}[htb]
\epsfig{file=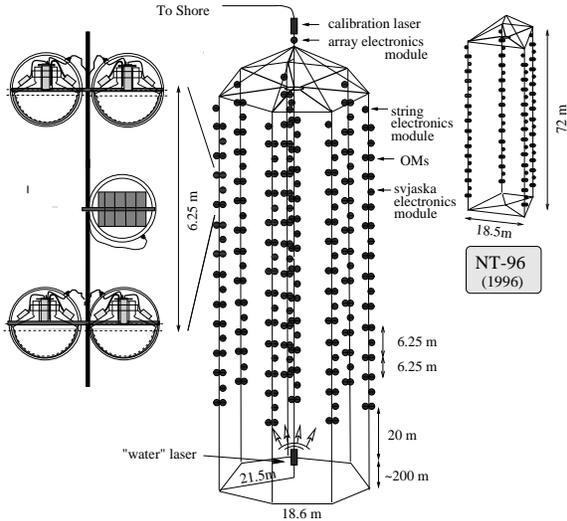,width=7.5cm}
\vspace{-0.8cm}
\caption{
Schematic view of the Baikal Telescope {\it NT-200}. 
The expansion left-hand shows 2 pairs of
optical modules ("svjaska") with the svjaska
electronics module, which houses
parts of the read-out and control electronics.
Top right, the 1996 array {\it NT-96} is sketched.            
}
\vspace{-6mm}
\label{fig1}
\end{figure}

%
%
A {\it muon-trigger}
is formed by the requirement of \mbox{$\geq N$ {\it hits}}
(with {\it hit} referring to a channel) within \mbox{500 ns}.
$N$ is typically set to 
\mbox{3 or 4.} For  such  events, amplitude and time of all fired
channels are digitized and sent to shore. 
A separate {\em monopole trigger} system searches for clusters of
sequential hits in individual channels which are
characteristic for the passage of slowly moving, bright
objects like GUT monopoles.

The main challenge of large underwater neutrino telescopes
is the identification of extraterrestrial neutrinos of
high energy. In this paper we present results of a search for
neutrinos with $E_{\nu}>10 \,$TeV obtained with the deep underwater
neutrino telescope {\it NT-96} at Lake Baikal \cite{APP2,JF}. 

\section{Search strategy and the limits on the diffuse neutrino flux}

The used search strategy for high energy neutrinos relies
on the detection of the Cherenkov 
light emitted by the electro-magnetic and (or) hadronic
particle cascades and high energy muons
produced at the neutrino interaction
vertex in a large volume around the neutrino telescope.

We select
events with high multiplicity of hit channels 
corresponding to bright cascades. The 
volume considered for generation of cascades is essentially
{\it below} the geometrical volume of {\it NT-96.}
A cut is applied
which accepts only time patterns corresponding to upward
traveling light signals (see below). 

Neutrinos produce showers and high energy muons through 
CC-interactions

\begin{equation}
\nu_l(\bar{\nu_l}) + N \stackrel{CC}{\longrightarrow} l^-(l^+) + 
\mbox{hadrons},
\end{equation}
through NC-interactions

\begin{equation}
\nu_l(\bar{\nu_l}) + N \stackrel{NC}{\longrightarrow} 
\nu_l(\bar{\nu_l}) + \mbox{hadrons},
\end{equation}
where $l=e$ or $\mu$, and through resonance production \cite{Glash,Ber1,Gandi}

\begin{equation}
\bar{\nu_e} + e^- \rightarrow W^- \rightarrow \mbox{anything},
\end{equation}
\noindent
with the resonant neutrino energy  
$E_0=M^{2}_w/2m_e=6.3\cdot 10^6 \,$GeV 
and cross section $5.02\cdot 10^{-31}$cm$^2$.

Within the first 70 days of effective data taking, $8.4 \cdot 10^7$ events
with $N_{hit} \ge 4$ have been selected. 

For this analysis we used events with $\ge$4 hits along at least one
of all hit strings. The time difference between any two channels
on the same string was required to obey the condition:

\begin{equation}
\mid(t_i-t_j)-z_{ij}/c\mid<a\cdot z_{ij} + 2\delta, \,\,\, (i<j).
\end{equation}
The $t_i, \, t_j$ are the arrival times at channels $i,j$, and
$z_{ij}$ is their vertical distance. 
$\delta=5$ nsec 
accounts for the timing error and $a=1 \,$ nsec/m. 

8608 events survive the selection criterion (4).
The highest multiplicity of hit 
channels (one event) is $N_{hit}=24$.

Since no events with $N_{hit}>24$ are found in our data we can derive
upper limits on the flux of high energy neutrinos which produce 
events with multiplicity 

\begin{equation}
N_{hit}>25.
\end{equation}

The shape of the neutrino spectrum was assumed to behave like 
$E^{-2}$ as typically expected for Fermi acceleration.
In this case, 90\% of expected events would be produced by  neutrinos
from the energy range $10^4 \div 10^7$GeV.
Comparing the calculated rates with the upper limit to the 
actual number of events, 2.3 for 90\% CL
we obtain the following upper
limit to the diffuse neutrino flux:

\begin{equation}
\frac{d\Phi_{\nu}}{dE}E^2<1.4\cdot10^{-5} 
\mbox{cm}^{-2}\mbox{s}^{-1}\mbox{sr}^{-1}\mbox{GeV}.
\end{equation}

\begin{figure}[htb]
\epsfig{file=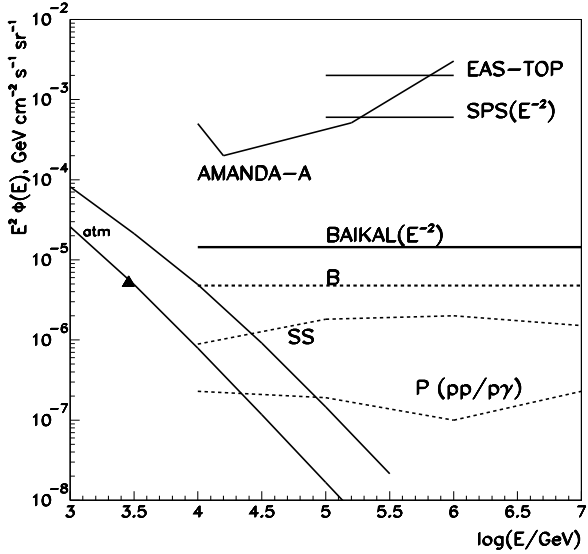,width=8.5cm,height=8.0cm}
\caption { 
Upper limits to the differential flux of high 
energy neutrinos obtained by 
different experiments as well as upper bounds for  
neutrino fluxes from a number of different models. 
The triangle denotes the FREJUS limit.
\vspace{-1.cm}
}
\label{fig5}
\end{figure}

Fig.2 shows the upper limits to the diffuse high energy neutrino
fluxes obtained by BAIKAL (this work), SPS-DUMAND \cite{DUMAND}, AMANDA \cite{AMANDA}, 
EAS-TOP \cite{EAS} and FREJUS \cite{FREJUS} 
(triangle) experiments as well as
model independent upper limit obtained by V.Berezinsky \cite{Ber3}
(curve labelled B)
(with the energy density of the diffuse X- and gamma-radiation 
$\omega_x \leq 2 \cdot 10^{-6}$ eV cm$^{-3}$
as follows from EGRET data \cite{EGRET}) and 
the atmospheric neutrino fluxes \cite{LIP} 
from horizontal  and vertical directions (upper and lower curves,
respectively).
Also, predictions from Stecker and Salamon model \cite{SS} 
(curve labelled SS) and
Protheroe model \cite{P} (curve labelled P) for diffuse neutrino fluxes from quasar cores
and blazar jets are shown in Fig.2.

For the resonant process (3) 
our 90\% CL limit at the W resonance energy is:

\begin{equation}
\frac{d\Phi_{\bar{\nu}}}{dE_{\bar{\nu}}} \leq 3.6 \times 
10^{-18} 
\mbox{cm}^{-2}\mbox{s}^{-1}\mbox{sr}^{-1}\mbox{GeV}^{-1}.
\end{equation}

The limit (6) obtained for the diffuse neutrino flux 
is of the same order as the limit announced
by FREJUS \cite{FREJUS} but extends to much
higher energies. 
We expect that analysis of 3 years
data taking with {\it NT-200} would allow us to lower 
this limit by another order of magnitude.

\bigskip

{\it This work was supported by the Russian Ministry of Research,the German 
Ministry of Education and Research and the Russian Fund of Fundamental 
Research \mbox{( grants }} \mbox{\sf 99-02-18373a}, \mbox{\sf 97-02-17935}, 
\mbox{\sf 99-02-31006} {\it and} \mbox{\sf 97-15-96589}),
{\it and by the Russian Federal Program ``Integration'' (project no.} 346).

\end{document}